\documentclass[11pt]{article}
\usepackage{moriond,epsfig}
\usepackage{xspace, relsize, amsmath}

\def\Journal#1#2#3#4{{#1} {\bf #2}, #3 (#4)}

\def\NIMA{{\em Nucl. Instrum. Methods} A}

\def\PLB{{\em Phys. Lett.}  B}
\def\PRL{\em Phys. Rev. Lett.}
\def\PRD{{\em Phys. Rev.} D}

\def\be{\begin{equation}}
\def\ee{\end{equation}}
\def\bea{\begin{eqnarray}}
\def\eea{\end{eqnarray}}

\def\babar{\mbox{\slshape B\kern-0.1em{\smaller A}\kern-0.1em
    B\kern-0.1em{\smaller A\kern-0.2em R}}}
\def\B   {\ensuremath{B}\xspace}
\def\Bz  {\ensuremath{B^0}\xspace}
\def\Bc  {\ensuremath{B^+}\xspace}
\def\BB  {\ensuremath{B\overline{B}}\xspace}
\def\bll {\ensuremath{\B^0\rightarrow\ell^+\ell^-}\xspace}
\def\bln {\ensuremath{\B^+\rightarrow\ell^+\nu}\xspace}
\def\blt {\ensuremath{\B^0\rightarrow\ell^+\tau^-}\xspace}
\def\bknn {\ensuremath{\B\rightarrow K\nu\overline\nu}\xspace}
\def\ifb  {fb\ensuremath{^{-1}}\xspace}
\def\hjCM {CM\xspace}
\def\mes {\ensuremath{m_{ES}}\xspace}
\def\dE {\ensuremath{\Delta E}\xspace}
\def\Fisher {\ensuremath{\mathcal{F}}\xspace}
\newcommand{\gev}{\ensuremath{\mathrm{\,Ge\kern -0.1em V}}\xspace}
\newcommand{\mev}{\ensuremath{\mathrm{\,Me\kern -0.1em V}}\xspace}
\newcommand{\gevc}{\ensuremath{{\mathrm{\,Ge\kern -0.1em V\!/}c}}\xspace}
\newcommand{\mevc}{\ensuremath{{\mathrm{\,Me\kern -0.1em V\!/}c}}\xspace}
\newcommand{\gevcc}{\ensuremath{{\mathrm{\,Ge\kern -0.1em V\!/}c^2}}\xspace}
\newcommand{\mevcc}{\ensuremath{{\mathrm{\,Me\kern -0.1em V\!/}c^2}}\xspace}

\newcommand{\splot}{\ensuremath{sPlot}\xspace}

\def\Btag{\ensuremath{B_{tag}}\xspace}

\def\etal{{\it et al.}}

\begin{document}
\vspace*{4cm}
\title{New Results on Leptonic \B meson decays at \babar}

\author{ Kim Hojeong}

\address{Stanford Linear Accelerator Center, 2575 Sand Hill Road, Menlo Park, CA, 94025, USA}

\maketitle\abstracts{
We present selected new results on leptonic \B meson decays from the \babar\ experiment: searches for the decays \bll, \bln and \blt, and \bknn, where~$\ell=e$~or~$\mu$~\cite{bib:conjugation}. We observe no evidence for these decays and set upper limits on their branching fractions.}

\section{Introduction}
Leptonic \B meson decays provide an important tool to investigate the Standard Model (SM) and physics beyond the SM. They are highly suppressed in the SM, because they involve a $b\rightarrow d$ transition, require an internal quark annihilation, and there are also helicity suppression for \bll and \bln modes, and because the flavor-changing neutral-currents are forbidden at the tree level for \bknn mode. The decay rates can be enhanced or reduced when heavy virtual particles like Higgs or super-symmetric~\cite{bib:SUSY} (SUSY) particles replace the W boson or show up at higher orders in loop diagrams. Constraints on these decays can provide information on important
SM parameters, such as \B meson decay constant. They have identifiable final states with low multipliticy, but they are mostly below our sensitivity. These decay modes will play an important role at the future colliders, such as a Super-B factory, ILC, and LHC (for muon modes).\\

The analyses described in this paper use data recorded with the \babar\ detector at the PEP-II asymmetric energy $e^+e^-$ storage rings. A detailed description of the \babar\ detector can be found elsewhere~\cite{bib:babar}. A full \babar\ Monte Carlo (MC) simulation using \texttt{GEANT4}~\cite{bib:geant4} is used to evaluate signal efficiencies and to identify and study background sources.

\section{\bll}

The leptonic decays \bll are studied using  $383.6\times 10^{6}$ \BB events. The SM prediction on the branching fractions (BFs) are $1.9\times 10^{-15}(8.0\times 10^{-11})$ for the $e^+e^-(\mu^+\mu^-)$ mode, and the $\Bz\rightarrow e^\pm\mu^\mp$ decay is forbidden. The best upper limits (UL) on the BFs have been set at the order of $10^{-8}$ by the \babar~\cite{bib:bllBaBar} experiment for $e^+e^-$ and $e^\pm\mu^\mp$ modes using 111\ifb, and by CDF~\cite{bib:bllCDF} experiment for $\mu^+\mu^-$ mode with 2\ifb.\\

The \Bz candidate is reconstructed by combining two oppositely charged tracks originating from a common vertex. We use two kinematic quantities: $\mes=\sqrt{(E_{\text{beam}}^*)^2-(\Sigma_i{\bf p}_i^*)^2}$ and $\dE=\sum_i\sqrt{m_i^2+({\bf p}_i^*)^2}-E_{\text{beam}}^*$, where $E_{\text{beam}}^*$ is the beam energy in the \hjCM frame, ${\bf p}_i^*$ and $m_i$ are the momenta in the \hjCM frame and the masses of the daughter particles $i$ of \B meson. $E_{\text{beam}}^*$ is used instead of the measured \B meson energy in the \hjCM frame because $E_{\text{beam}}^*$ is more precisely known. For correctly reconstructed \Bz mesons, the \mes distribution has a maximum at the \Bz mass with a standard deviation of about 2.5\mevcc and the \dE distribution has a maximum near zero with a standard deviation of about 25\mev.\\
 
Stringent requirements on particle identification~\cite{bib:PID} are made to reduce the contamination from misidentified hadrons and leptons. We retain about 93\% (73\%) of the electrons (muons), with a misidentification rate for pions of less than about 0.1\% (3\%). The main background are continuum processes where $e^+e^-\rightarrow f\overline{f}$, ($f$ = $u,d,s,c,\tau$). A Fisher discriminant~\cite{bib:fisher} (\Fisher) is constructed, using their different event topology with respect to that of the signal events. \\

A maximum likelihood (ML) fit is performed based on the variables \mes, \dE and \Fisher. The results are summarized in Table~\ref{tab:llresult}. The event and background \splot~\cite{bib:splot} distributions are shown in Figure~\ref{fig:llfit}. Using a Bayesian approach, a 90\% confidence level (CL) UL on the BF is calculated. The systematic uncertainties are included as a Gaussian into the likelihood calculation.

\begin{table}[t]
\caption{Result of \bll analysis. Efficiency ($\epsilon$), number of signal events ($N_{sig}$) from ML fit, and 90\% confidence level upper limit on the branching fraction (UL(BF)) for the three leptonic decays $\Bz\rightarrow e^+e^-$, $\Bz\rightarrow \mu^+\mu^-$, and $\Bz\rightarrow e^\pm\mu^\mp$ are shown. \label{tab:llresult}}
\vspace{0.4cm}
\begin{center}
\begin{tabular}{|c|c|c|c|}
\hline
&$\epsilon$ (\%) & $N_{sig}$ & UL(BF)$\times 10^{-8}$\\ \hline
$\Bz\rightarrow e^+e^-$ & $16.6\pm 0.3$ & $0.6\pm 2.1$ & $11.3$\\
$\Bz\rightarrow \mu^+\mu^-$ & $15.7\pm 0.2$ & $-4.9\pm 1.4$ & $5.2$\\
$\Bz\rightarrow e^\pm\mu^\mp$ & $17.1\pm 0.2$ & $1.1\pm 1.8$ & $9.2$\\ \hline
\end{tabular}
\end{center}
\end{table}

\begin{figure}
\begin{center}
\includegraphics[height=4cm]{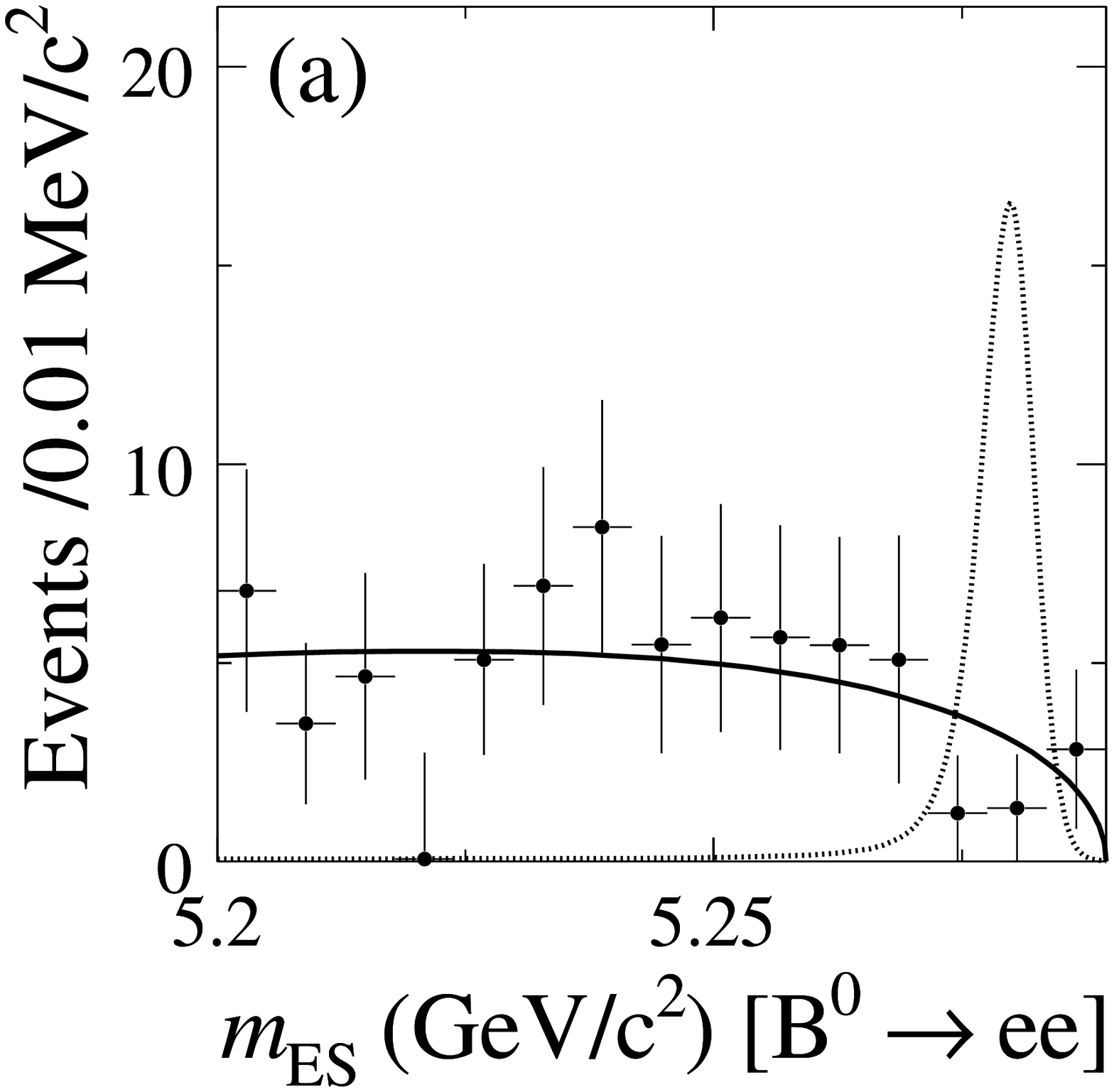}\hspace{0.1 in}
\includegraphics[height=4cm]{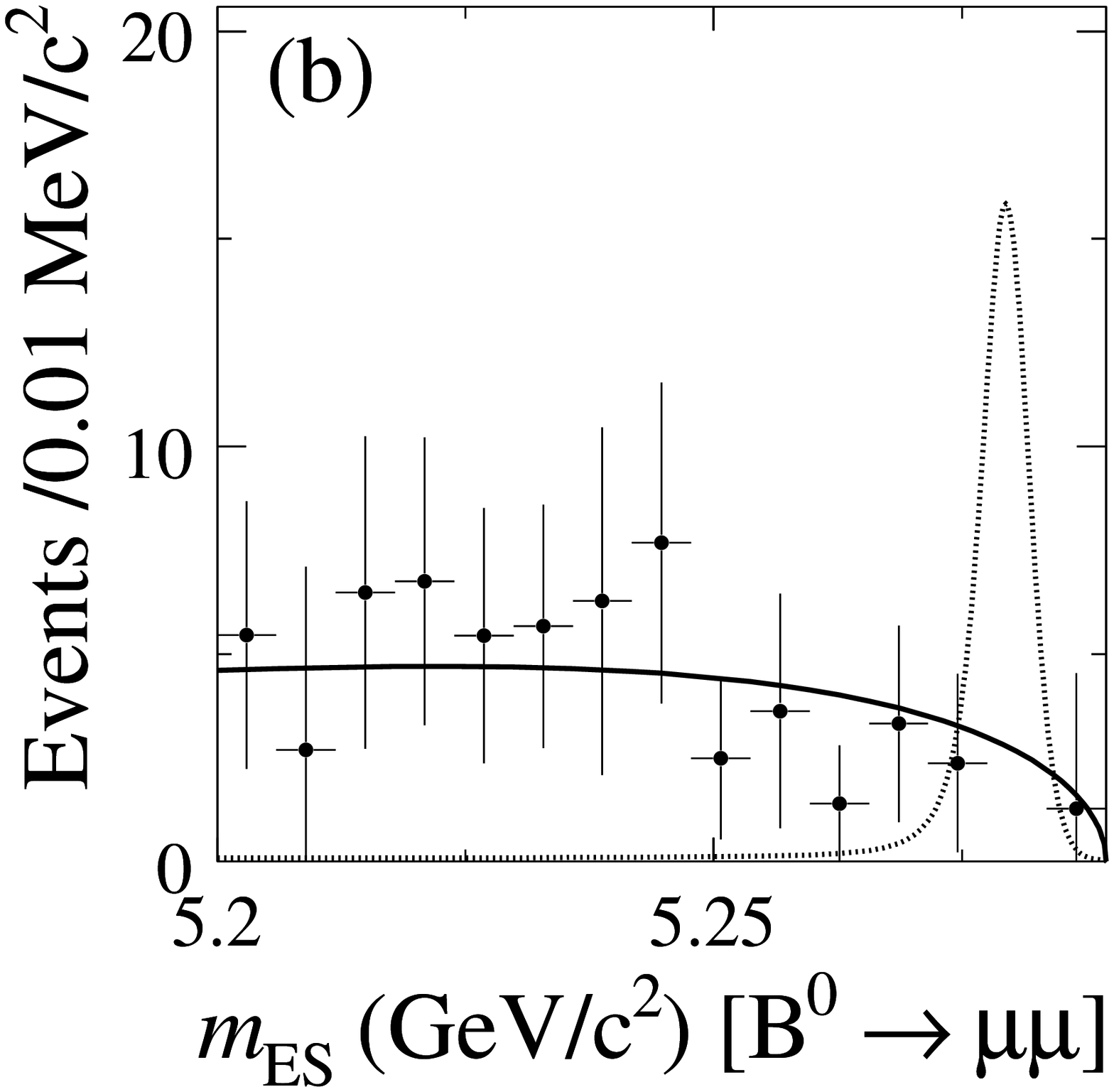}\hspace{0.1 in}
\includegraphics[height=4cm]{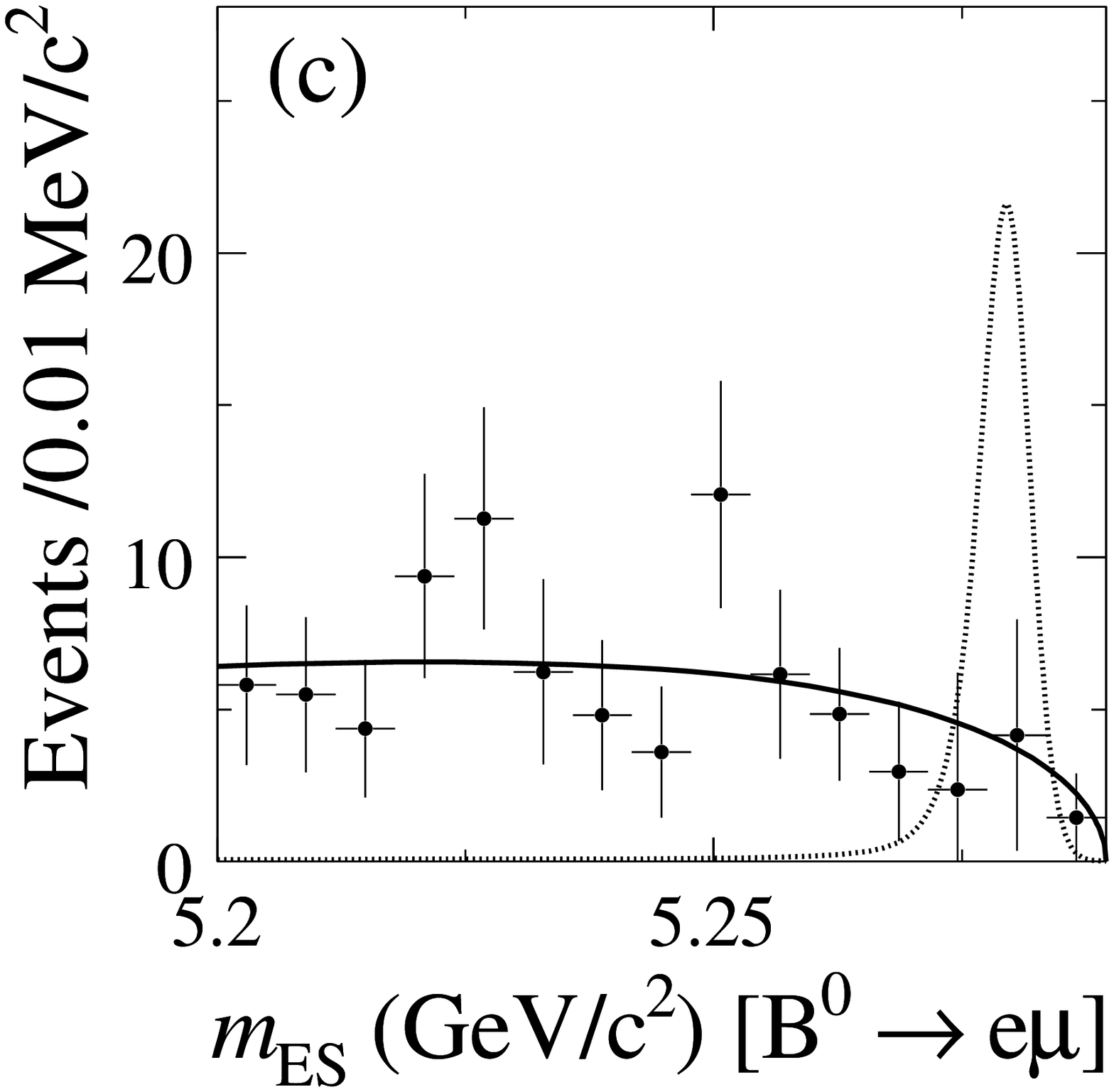}\\
\includegraphics[height=4cm]{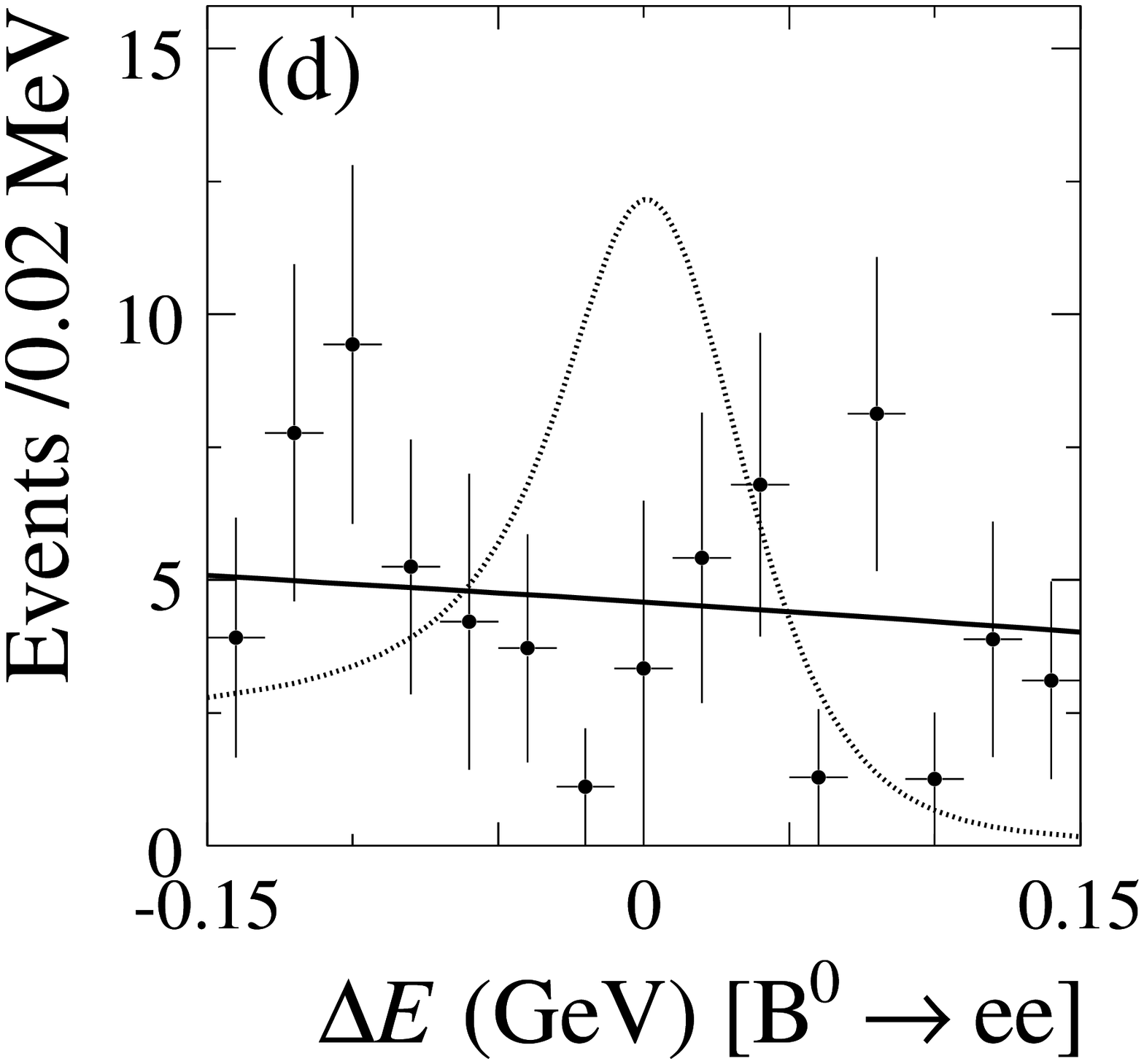}\hspace{0.1 in}
\includegraphics[height=4cm]{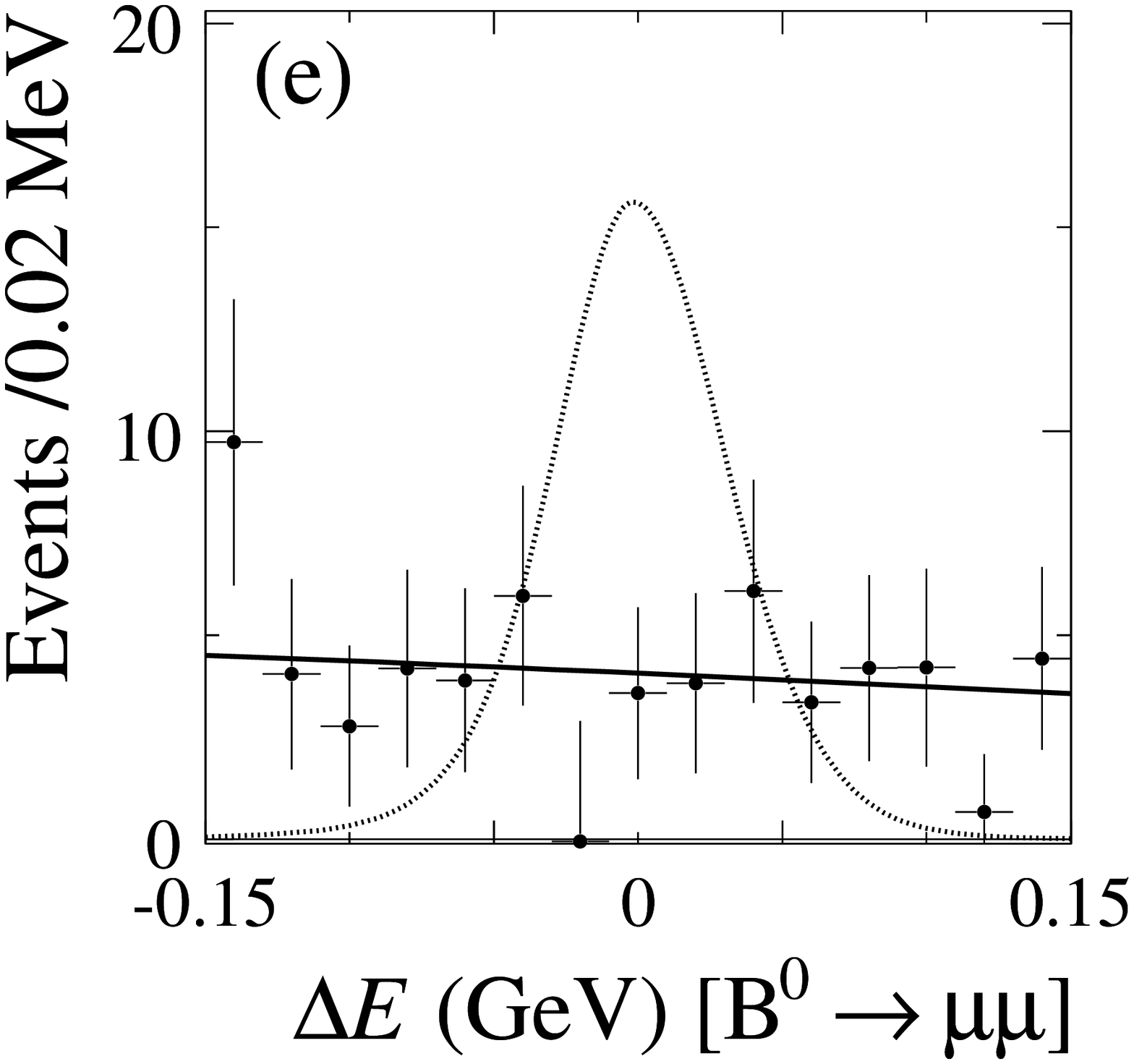}\hspace{0.1 in}
\includegraphics[height=4cm]{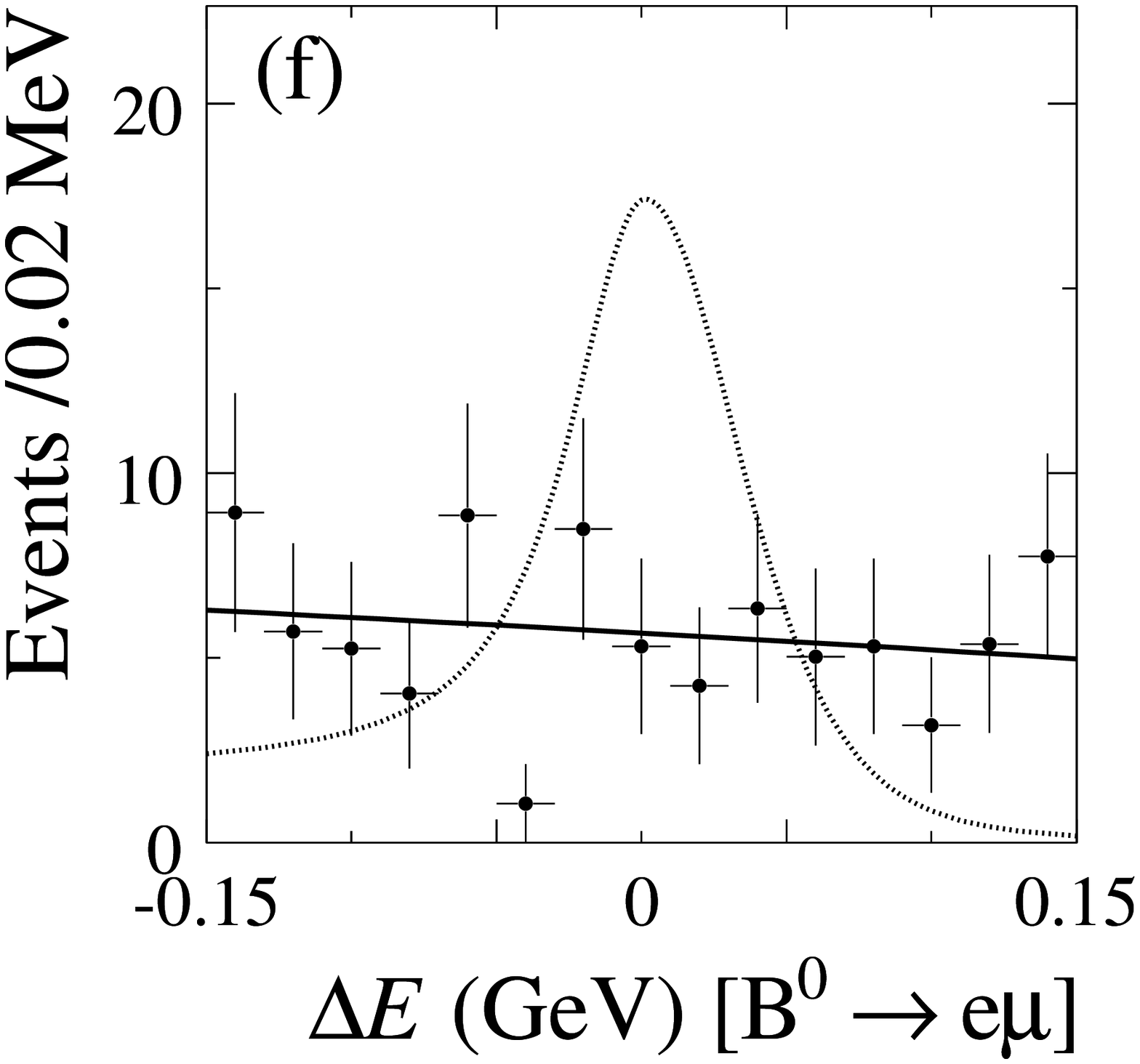}\\
\includegraphics[height=4cm]{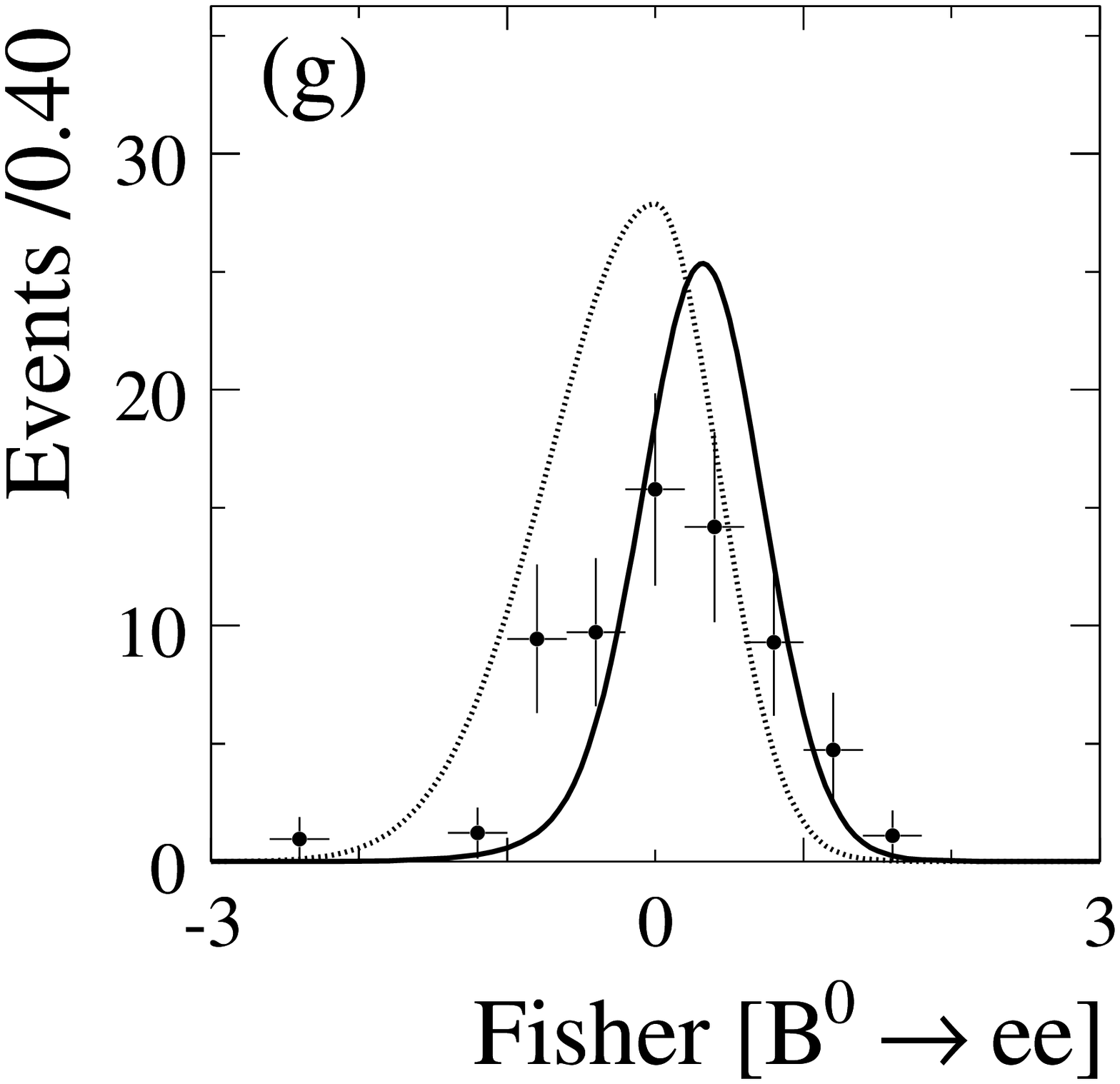}\hspace{0.1 in}
\includegraphics[height=4cm]{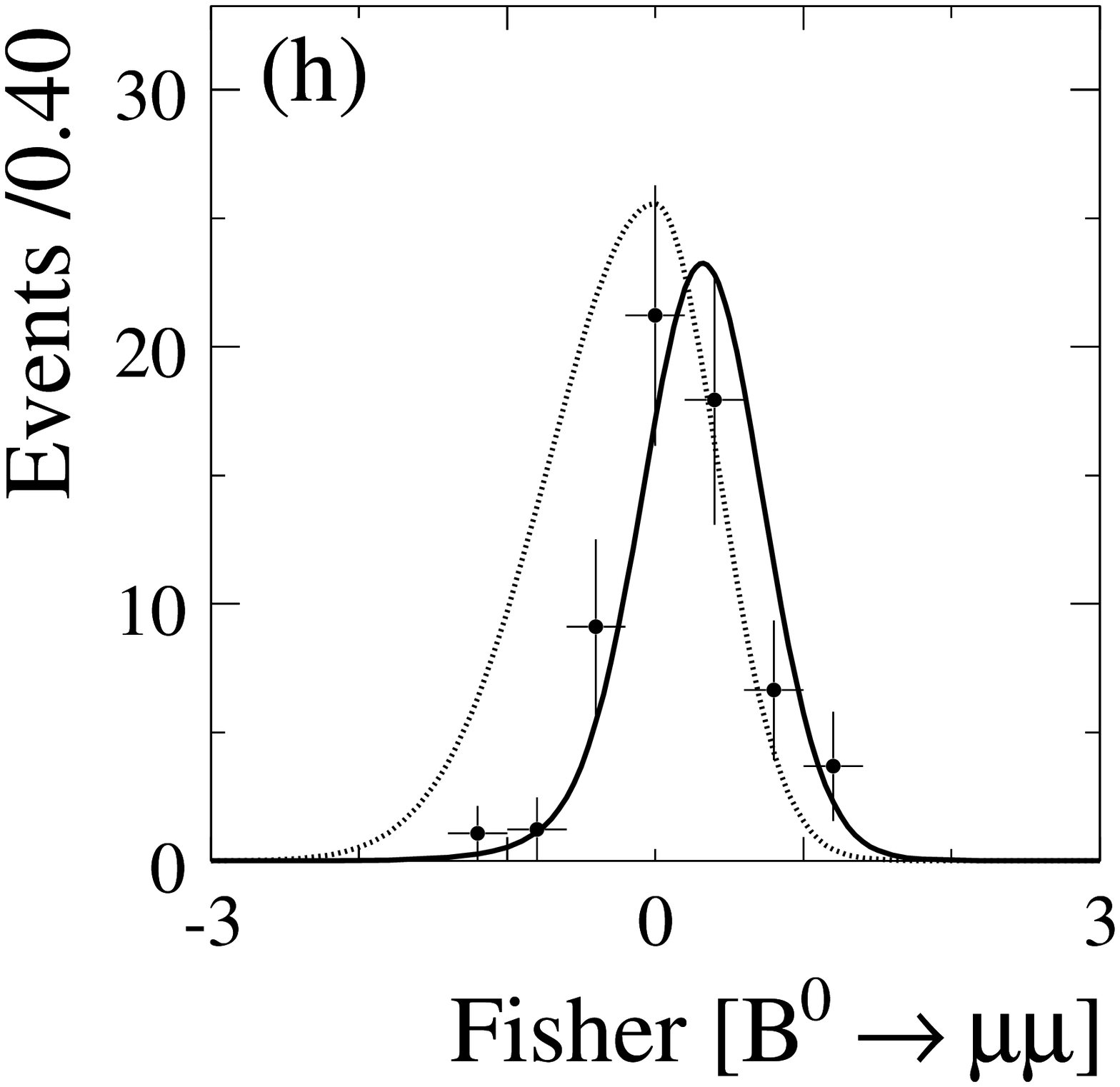}\hspace{0.1 in}
\includegraphics[height=4cm]{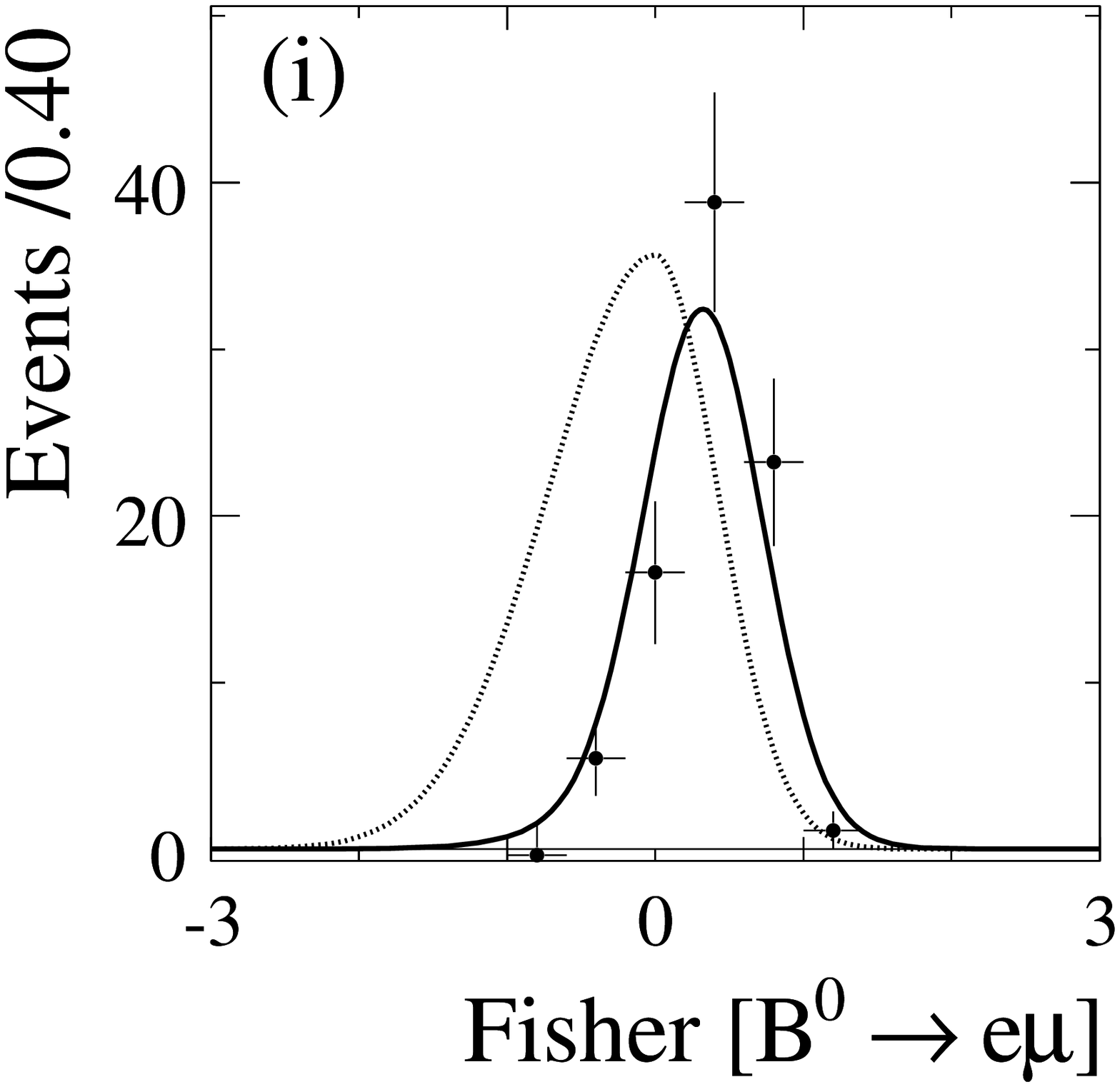}\\

\caption{The distributions of events in \mes (a,b,c), \dE (d,e,f) and \Fisher (g,h,i) for $\Bz\rightarrow e^+e^-$ (left), $\Bz\rightarrow \mu^+\mu^-$ (middle), $\Bz\rightarrow e^\pm\mu^\mp$ (right) are shown. The points with error bars are data. The overlaid solid curve in each plot is the background \splot distribution obtained by maximizing the likelihood not using the information from the corresponding component.  The dotted line, representing the signal probability density function with an arbitrary scaling, indicates where the signal is expected.}
\label{fig:llfit}
\end{center}
\end{figure}

\section{\bln and \blt}
We present searches for the decays \bln and the lepton flavor violating decays \blt, where $\ell = e$ or $\mu$ using 378$\times 10^6$ \BB events. The SM predictions of the BFs are of the order of $10^{-11} (10^{-7})$ for $\Bc\rightarrow e^+\nu$ $(\Bc\rightarrow\mu^+\nu)$, and \blt modes are forbidden. The UL on the BFs have been measured by \babar~\cite{bib:bltBabar}, Belle~\cite{bib:bltBelle}, and CLEO~\cite{bib:bltCLEO}. The best published limits are from Belle for \bln, at the order of $10^{-6}$ with 253\ifb, and CLEO for \blt, at the order of $10^{-4\sim -5}$ with $9.6\times 10^{6}$ \BB events.\\

We fully reconstruct one of the two \B mesons (\Btag) in the event: $\B_{tag}\rightarrow D^{(*)}X_{had}$, $X_{had}$ decays in combinations of $K$'s and $\pi$'s. This method has not been used for searches for these modes. To suppress the continuum backgrounds, we use their different event topologies with respect to that of the signal events. After all selection criteria are applied, it results in a yield of approximately 2500 (2000) correctly reconstructed \Bc (\Bz) candidates per \ifb of data. This hadronic tagging method yields lower statistics than other methods but it provides an almost background-free environment.\\

All particles not used in the \Btag reconstruction are included in the reconstruction of the signal \B meson. From the two-body kinematics, we expect a mono-energetic lepton in the signal \B rest frame: lepton momentum ($p^*$) of 2.64 (2.34) \gevc for the \bln (\blt) modes.\\

We reconstruct $\tau$ in the following modes: $e^-\overline\nu_e\nu_\tau$, $\mu^-\overline\nu_\mu\nu_\tau$, $\pi^-\nu_\tau$,  $\pi^-\pi^0\nu_\tau$, $\pi^-\pi^0\pi^0\nu_\tau$, and $\pi^-\pi^-\pi^+\nu_\tau$. The second highest momentum track in the event excluding the \Btag daughters is assumed to be a $\tau$ daughter, and is required to have a charge opposite to the primary signal lepton. \\

The signal yields are extracted from unbinned ML fits to the signal lepton momentum distributions, as measured in the signal \B rest frame. The fits are restricted to the ranges in $p^*$ shown in Fig.~\ref{fig:blnfit}. Using a Bayesian approach, a 90\% CL UL on the BF is determined. The dominant systematic uncertainties are due to the fitting procedure and the determination of \Btag efficiencies. The total uncertainty is between 10 and 16\% depending on the modes. The uncertainties are incorporated into the final results by varying the BF assumption by its uncertainty when integrating likelihood for the 90\% CL UL. The results are summarized in Table~\ref{tab:blnresult}.

\begin{figure}
\begin{center}
\includegraphics[height=4cm]{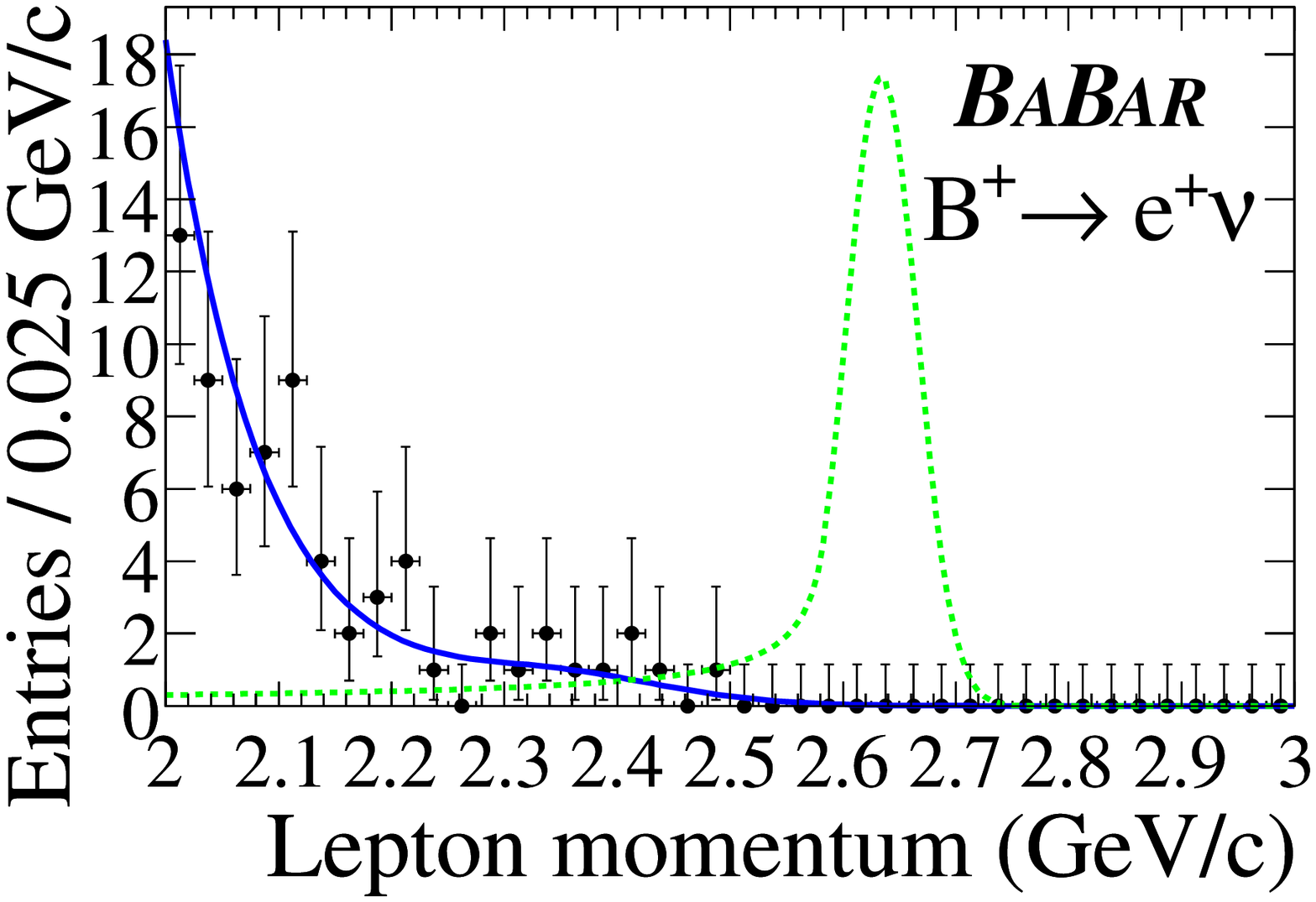}\hspace{0.1 in}
\includegraphics[height=4cm]{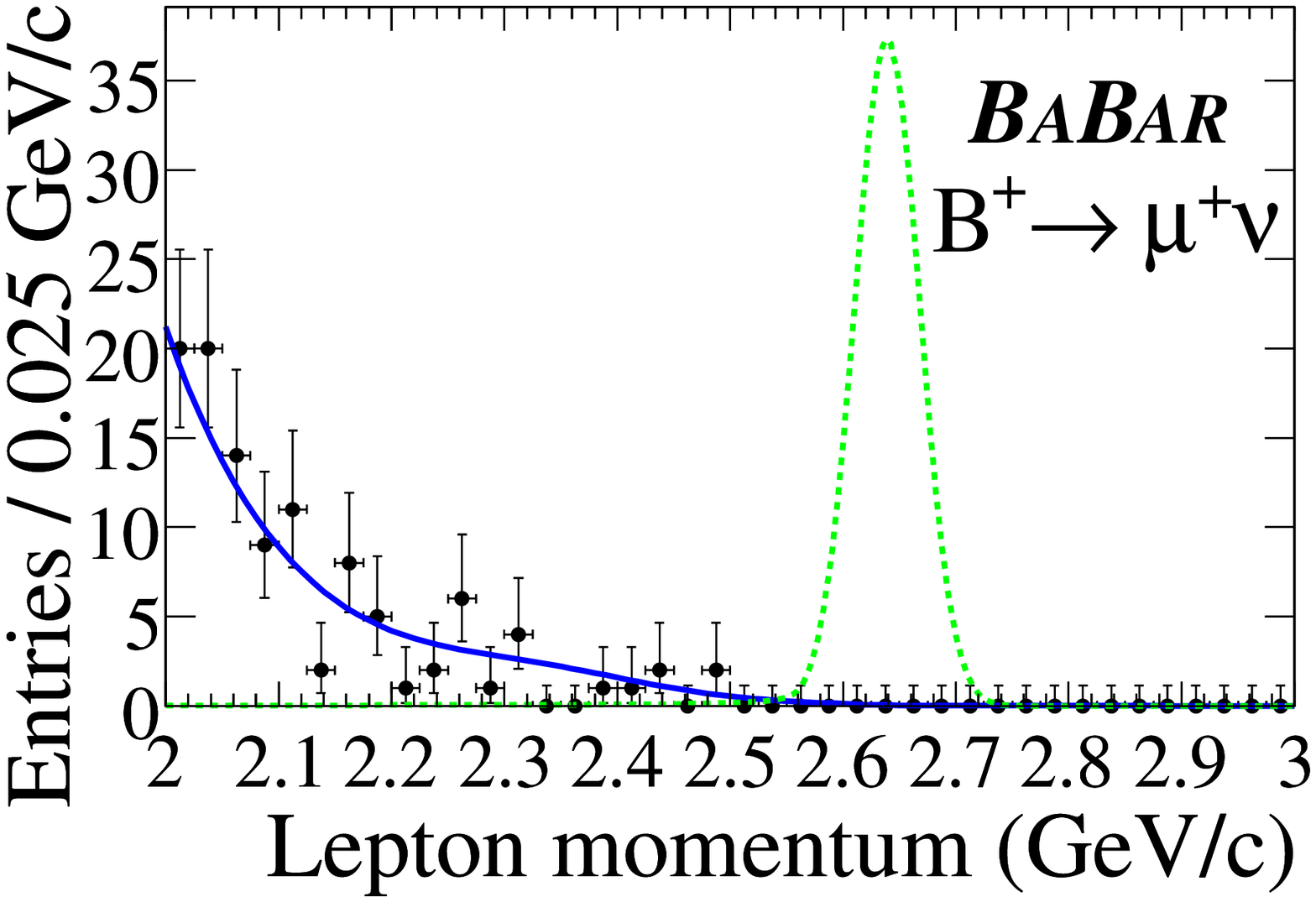}\\
\includegraphics[height=4cm]{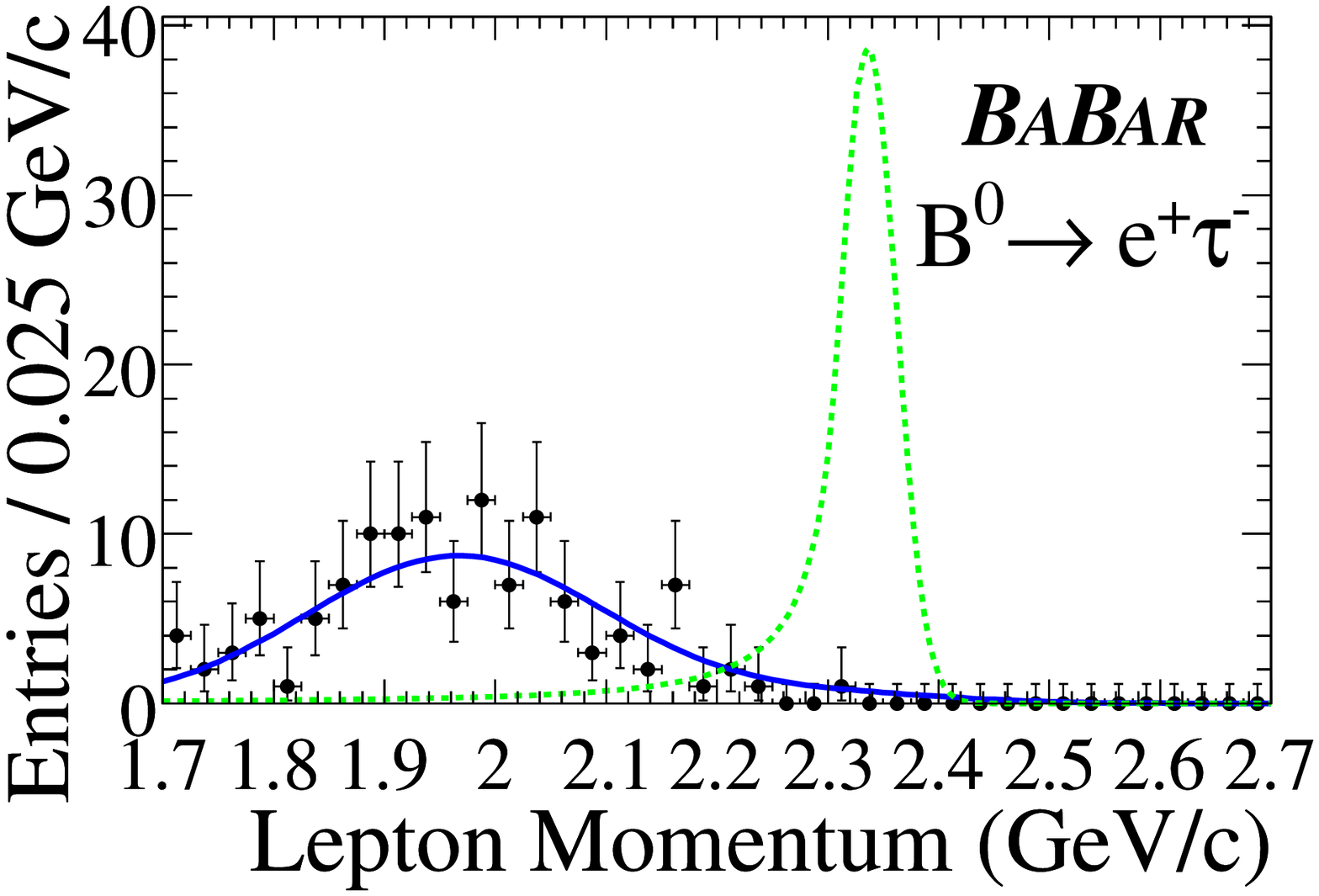}\hspace{0.1 in}
\includegraphics[height=4cm]{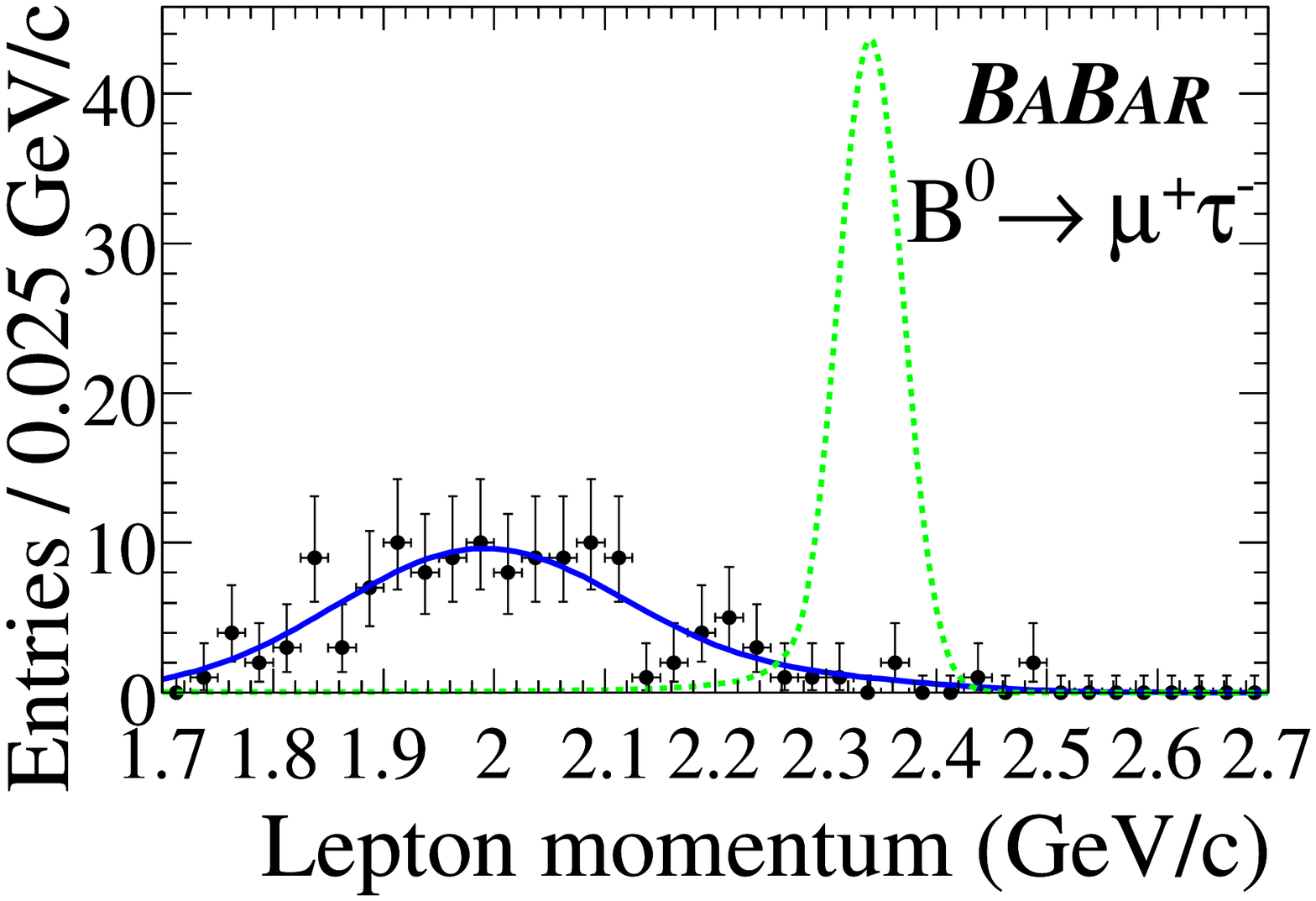}
\caption{The unbinned maximum likelihood fits and the distributions of the lepton momentum for \bln and \blt analyses. The points with error bars are data, the solid line represents the ML fit. The dashed line, representing the signal probability density function with an arbitrary scaling, indicates where the signal is expected.}
\label{fig:blnfit}
\end{center}
\end{figure}

\begin{table}[t]
\caption{Result of \bln and \blt anayses. The efficiency ($\epsilon$), number of signal events ($N_{sig}$) and 90\% CL UL on the BF (UL(BF)) for the decay modes are shown. \label{tab:blnresult}}
\vspace{0.4cm}
\begin{center}
\begin{tabular}{|c|c|c|c|}
\hline
&$\epsilon$ $(\times 10^{-5})$ & $N_{sig}$ & UL(BF)$\times 10^{-6}$\\ \hline
$\Bc\rightarrow e^+\nu$ & $135\pm 4$ & $-0.07\pm 0.03$ & $5.2$\\
$\Bc\rightarrow\mu^+\nu$ & $120\pm 4$ & $-0.11\pm 0.05$ & $5.6$\\
$\Bz\rightarrow e^+\tau^-$ & $32\pm 2$ & $0.02\pm 0.01$ & $28$\\
$\Bz\rightarrow\mu^+\tau^-$ & $27\pm 2$ & $0.01\pm 0.01$ & $22$\\ \hline
\end{tabular}
\end{center}
\end{table}

\section{\bknn}
The \bknn decays are studied using 319 \ifb of data. The SM prediction of this mode~\cite{bib:bknnSM} is $(3.8\pm 1.2)\times 10^{-6}$ and the best published UL is at $1.4\times 10^{-5}$ from Belle~\cite{bib:bknnBelle} with $535\times 10^{6}$ \BB events.\\

We reconstruct one of the two \B mesons in the event, where it decays semileptonically: $\Bc\rightarrow D^{(*)0}\ell^+\nu$. Compared to hadronic tagging method used in in \bln and \blt analyses, this semileptonic tagging method yields higher statistics with more background. \\

A multivariate classifier, the Random Forest (RF) tool from StatPatternRecognition~\cite{bib:SPR} is used to optimize signal separation from background. Several regions of the parameter space (terminal leaf size, maximum number of input variables randomly selected for decision splits) are explored with the RF classifier. We use the Punzi Figure of Merit~\cite{bib:punzi}, $S/(N_{\sigma}/2 + \sqrt{b})$, where $s$ is signal, $b$ is background and $N_{\sigma}$ is the sigma level of discovery (we take $N_\sigma=3$), and found the optimal Punzi Figure of Merit with a terminal leaf size of 35 events, after growing 100 decision trees, and sampling on at most 20 variables. The variables include number of tracks in the event (excluding tracks from the \Btag reconstruction), transverse momentum of tracks, event topology variables, missing energy in the event, total energy in the event, total energy deposit in the detector that are not associated with any charged or neutral particles.\\

The signal box is defined in the 2-dimensional space of $D^0$ mass and the RF output, which is blinded until we finish with all selections and estimations. The RF output ranges between 0 and 1. The signal box is RF output bigger than 0.82 and near $D^0$ mass peak which varies depends on the $D$ modes. We estimate the background level in the signal box using MC events as well as data  outside of the signal box.\\

While $30.71\pm 10.71$ events are expected 38 events are observed as shown in Figure~\ref{fig:knnresult}. The systematic uncertainties, which are estimated using double tag events, in where both \B mesons decay semileptonically, are incorporated in the UL BF calculation. We set 90\% UL BF at $4.2\times 10^{-5}$, using a modified frequentist method~\cite{bib:barlow}.

\begin{figure}
\begin{center}
\includegraphics[height=6.5cm]{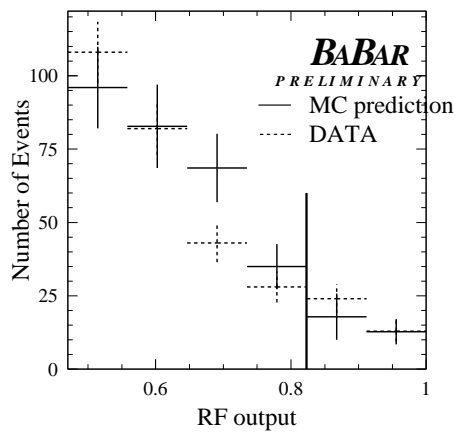}
\caption{The output of the Random Forest for \bknn analysis. The right side of the black line is the signal box. The dashed line is data and the solid line is expected background from MC.}
\label{fig:knnresult}
\end{center}
\end{figure}

\section{Summary}
New leptonic \B meson decays from \babar\ are presented: \bll, \bln, \blt and \bknn decays. We have not observed signal and set upper limits on all of these decays. With much more statistics from Super-B factory or ILC, exploiting the hadronic tagging method may be powerful. The leptonic \B meson decays will provide us important information on nature with more data.

\end{document}
